# Spatially separated superconductivity and enhanced charge-density-wave ordering in IrTe$_2$ nano-flake

Yanpeng Song[1], Fanqi Meng[1], Tianping Ying[1], Jun Deng[1,4], Junjie Wang[1,4], Xu Han[1], Qinghua Zhang[1], Yuan Huang[2*], Jian-gang Guo[1,3*], Xiaolong Chen[1,3,4*]

1. Beijing National Laboratory for Condensed Matter Physics, Institute of Physics, Chinese Academy of Sciences, Beijing 100190, China
2. Advanced Research Institute of Multidisciplinary Science, Beijing Institute of Technology, Beijing 100081, China
3. Songshan Lake Materials Laboratory, Dongguan, Guangdong 523808, China
4. University of Chinese Academy of Sciences, Beijing 100049, China

**ABSTRACT:** The interplay among various collective electronic states such as superconductivity (SC) and charge density wave (CDW) is of tremendous significance in low-dimensional transition metal dichalcogenides. Thus far, a consensus on the relationship between SC and CDW has not been established in IrTe$_2$, where either competing or collaboration pictures have been suggested in the bulk or thick flakes. Here, we use the state-of-art Au-assisted exfoliation method to overcome the obstacle of interlayer Te-Te bonding, cleaving the IrTe$_2$ down to monolayer for the first time. A striking feature revealed by angle-resolved polarized Raman spectroscopy (ARPRS) and electrical transport measurements is the concurrence of phase separation in one single piece of nano-flake, *i.e.* the superconducting (*P*-3*m*1) and CDW (*P*-3) area. In the pure SC area, the dimensional fluctuations completely suppress the CDW ordering and induce SC at 3.5 K. Meanwhile, the pure CDW area with much enhanced $T_{CDW}$ at 605 K (compared to $T_{CDW}$ = 280 K in bulk) is always accompanied by a unique wrinkle pattern. Detailed analyses suggest the local strain-induced bond breaking of the Te-Te dimer facilitates the emergence of the CDW order. Our works provide compelling evidence of competition between SC and CDW, highlighting the importance of microstructure in determining the ground states of IrTe$_2$.

The competition, coexistence, or cooperation of various collectively ordered electronic states has been a breeding ground for novel states of matter.[1,2] Charge density wave (CDW) order, one of the most studied correlated states, is a periodic modulation of the atoms and conduction electron densities in low-dimensional compounds. Strong electron-phonon coupling would open a small gap on the Fermi surface and induce a metal-insulator transition. As for superconductivity (SC), the strong electron-phonon interaction leads to condensations of paired-electrons within the BCS framework. In transition metal dichalcogenides (TMDC) of $Cu_xTiSe_2$,[3] $1T$-$TaS_2$[4,5] and $1T$-$TiSe_2$,[6,7,8] applying chemical doping or external pressure suppresses the CDW order and concomitantly induces the SC, establishing the competition picture between them. However, different scenarios like coexistence and cooperation have been reported as the samples are made into monolayers or ultrathin flakes of $1T$-$TaS_2$,[5] $2H$-$NbSe_2$,[9,10,11] and $1T$-$TiSe_2$.[12] The electron-phonon coupling[13] and dimensional fluctuations[14] are proposed to be responsible for the competing pictures.

Among many TMDCs, 1T-$IrTe_2$ is a unique one. This high-atomic number material is predicted as a type-II Dirac semimetal with a Dirac point slightly above the Fermi level.[15] $1T$-$IrTe_2$ crystallizes in a space group of $P$-$3m1$, where the quasi-two-dimensional $IrTe_2$ layer consists of edge-sharing $IrTe_6$ octahedral.[16] In the bulk crystal, a stripe CDW transition with $q = 1/5(1, 0, -1)$ occurs at $T_{CDW}$ ~ 280 K and the structural symmetry lowers to $P$-$1$[17], $P1$[18] or $C2/m$[19]. Subsequent experiments revealed that there are strong interlayer Te-Te covalent bonds, leading to unusual $Te^{1.5-}$ and $Ir^{3+}$ at 300 K. At temperatures below $T_{CDW}$, the Te-Te bonds partially break, simultaneously the $Ir^{4+}$-$Ir^{4+}$ dimers form.[20,21,22,23] The phase transition is more likely to be driven by the Ir $5d_{xy}$ orbital ordering instead of Fermi surface nesting.[21] Upon Pd/Pt doping, the CDW is destructed and then the SC emerges at 3 K.[16,24,25] However, applying physical pressure, chemical pressures, and in-plane strain enhance the $T_{CDW}$, namely the weaker Te-Te bonding state favors a stronger CDW ordering.[20,25,26] This unusual enhancement of $T_{CDW}$ is in sharp contrast to the behaviors in other abovementioned TMDCs. The special covalent character in the Ir-Ir and Te-Te dimers as well as charge disproportionation makes the origin of structural/CDW transition disputable.

Recently, the interplay of CDW and SC in thinned $IrTe_2$ crystals has attracted broad attention. Electrical transport measurements reveal that the CDW and SC coexist in samples with a thickness of ~75 nm[27]. However, $IrTe_2$ with micrometer thickness exhibits either SC or CDW ground state depending on the post-quenching rate, suggesting a competition scenario.[28] In the latest report, the SC is fully encompassed by the stripe CDW order with a collaborating relationship.[29] According to their data, the micro/macroscopic phase separation spontaneity emerges in the thicker samples. To clarify the relationship between SC and CDW, we have to find a suitable method to obtain a thinner sample and separate different phases.

In this work, we use the newly-developed Au-assisted mechanical exfoliation method to exfoliate the $IrTe_2$ crystal to a thickness of 10 nm or monolayer for the first time. We unambiguously distinguish two spatially separated superconducting and CDW areas in one single piece of nanoflake by the angle-resolved polarized Raman spectroscopy (ARPRS) and electrical transport measurements. In area #1, the structural symmetry is $P$-3$m$1 and the SC is observed at 3.5 K. However, in area #2, the CDW ordering lowers structural symmetry to $P$-3, more significantly, the $T_{CDW}$ is enhanced to 605 K, much higher than the 280 K in the bulk. We propose that dimensional fluctuations and increased $c/a$ ratio are responsible for the SC and unusual enhancement of CDW order, respectively. Based on these results, we reach a unified picture of the competing CDW ordering and SC in IrTe$_2$.

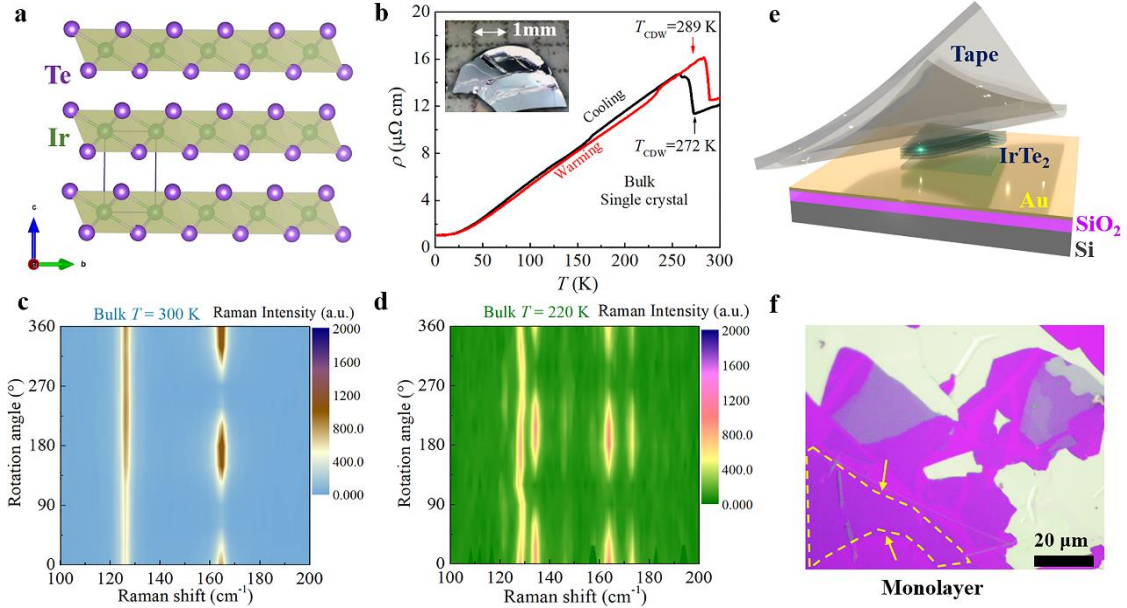

**Figure 1.** (a) Schematic structure of IrTe$_2$ at 300 K. (b) Temperature-dependence resistivity of bulk IrTe$_2$ from 2 K to 300 K. Inset is the optical image of IrTe$_2$ bulk crystal. (c-d) Angle-dependent Raman intensity of bulk IrTe$_2$ at 300 K and 220 K. (e) Illustration of Au-assisted exfoliation method. (f) Optical image of monolayer and few-layer IrTe$_2$.

Figure 1a shows the crystal structure of IrTe$_2$. There are interlayer Te-Te covalent bonds, i.e. dimerization of Te-Te, leading to relatively strong binding energy between IrTe$_2$ layers instead of van der Waals interaction.[30] The plate-like single crystal of IrTe$_2$ is shown in the inset of Fig. 1b. The temperature-dependent resistivity $\rho$(T) of IrTe$_2$ bulk crystal is plotted in Fig. 1b, where the jump and the hysteresis due to the CDW transition are observed in the range of 272 K and 289 K without any SC signal observed down to 2 K. In Fig. S1, the temperature-dependent Raman spectra exhibit the splitting of $E_g$ of 126 cm$^{-1}$ (in-plane stretching vibration of Te) and $A_{1g}$ of 164 cm$^{-1}$ (out-of-plane vibration of Te) at temperatures below $T_{CDW}$. Two peaks split into six peaks at 122, 128, 134, 145, 164 and 173 cm$^{-1}$, which are consistent with the reported feature.[31,32] The ARPRS of bulk IrTe$_2$ at 300 K and 220 K are measured for the first time to clarify the debated structural symmetry. The intensity ($I$) of a given Raman peak is proportional to the Raman tensor and scattering geometry,[33]

$$I \propto \sum_j |e_i \cdot R_j \cdot e_s|^2,$$

where $R_j$ is an arbitrary 3×3 Raman tensor, $e_i$ and $e_s$ are unit polarization vectors of the incident and scattered light, respectively. The angle between $e_i$ and $e_s$ is $\alpha$. The angle-dependent Raman intensity is plotted in Fig. 1c and d. According to the polarization configuration and selection rules of the space group $P$-$3m1$,[34] the intensities of $E_g$ is independent of α and $A_{1g}$ show a two-fold symmetry. Therefore, the peaks at 128 cm$^{-1}$ and 164 cm$^{-1}$ observed in various polarization configurations are assigned to the $E_g$ and $A_{1g}$ modes, respectively. The ARPRS below $T_{CDW}$ are plotted in Fig. S2. The peak at 128 cm$^{-1}$ observed in all polarization configurations is assigned as the $E_g$ mode. The peaks at 134, 145, and 173 cm$^{-1}$ show two-fold symmetry, which are the $A_g$ modes. According to the symmetry of the modes against α, we refer to that the space group below $T_{CDW}$ as $P$-3. We also use a specific Raman tensor of $C2/m$ to reproduce the experimental data,[34] and find that the calculated α-dependent intensity totally differs from the observed peak at 128 cm$^{-1}$. Moreover, as for $P$-1, there should be 21 Raman-active peaks $\Gamma_{Raman}$ =21 $A_g$ (xx, yy, zz, zy, xz, yz),[31] which are not consistent with our data. Thus, we rule out the possibility of $C2/m$ and $P$-1.

As mentioned above, the interlayer Te-Te is so strong that we have to use the newly-developed Au-assisted method[35] to exfoliate IrTe$_2$ nanoflakes. The deposited Au on the SiO$_2$/Si substrate can form Au-Te covalent bonding, which helps us to cleave the sample to thinner flakes. The typical process is illustrated in Fig. 1e and Note 1 in SI. By using this method, we successfully acquired a large-size sample with a planar area of 400 μm$^2$ and thickness less than 10 nm or even monolayer. The typical morphology of IrTe$_2$ thin flake is shown in Fig. 1f. Monolayer and few-layer IrTe$_2$ samples are clearly observed. In the Raman spectrum of monolayer IrTe$_2$, there is only $E_g$ peak as shown in Fig. S3. The lack of $A_{1g}$ peak is a general feature of monolayer which has been reported in other monolayer categories.[36]

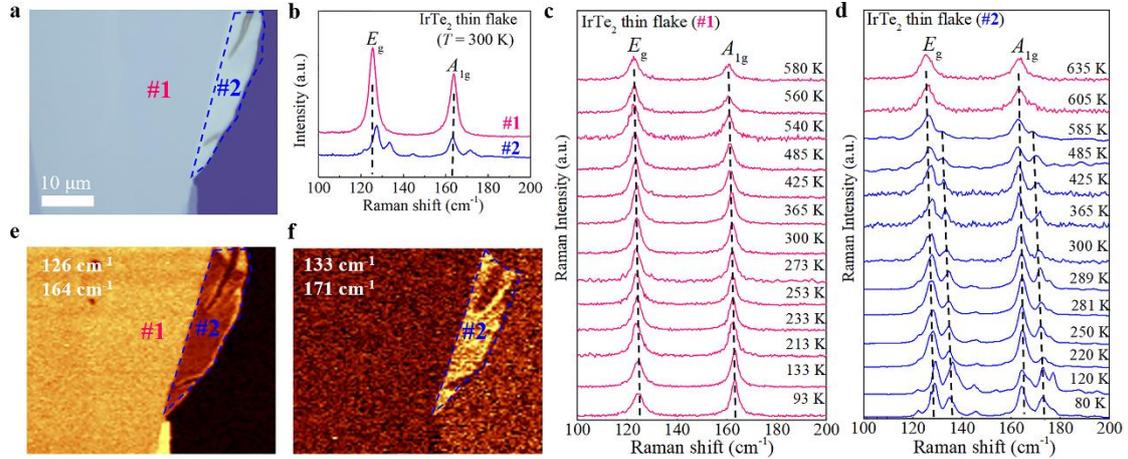

**Figure 2.** (a) Optical image of IrTe$_2$ nanoflake with wrinkles at the edge. The flat region is denoted as area #1. Area #2 with two wrinkles is marked by the blue dotted area. (b) Raman spectra of areas #1 and #2 collected at 300 K. (c-d) Temperature-dependent Raman spectra of areas #1 and #2. (e-f) Raman intensity mapping at 126/164 cm$^{-1}$ and 133/171 cm$^{-1}$.

Figure 2a presents an optical image of a typical exfoliated IrTe$_2$ nanoflake. We denote the flat region as area #1 and the blue-dotted area as area #2. There are two large strain-induced wrinkles at the boundary. This flat-wrinkle is also observed in other exfoliated samples (Fig. S4). Figure 2b shows the Raman spectra of two areas collected at 300 K.

For area #1, there are two sharp peaks at 126 cm$^{-1}$ and 164 cm$^{-1}$, identical to the spectrum measured at 300 K of bulk IrTe$_2$. On the other hand, in area #2, there are four additional peaks of 124, 133, 147, and 173 cm$^{-1}$ rasing from CDW ordering and the main peaks at 128 and 165 cm$^{-1}$. Besides, the frequencies of area #2 shift to a higher wavenumber than that of area #1 by 1-2 cm$^{-1}$ due to the shorter Te-Te bond length. A similar peak shift is observed in the Raman spectra of IrTe$_{2-x}$Se$_x$.[31]

We trace the temperature-dependent Raman spectra of two areas and plot them in Fig. 2c and 2d. In area #1, the $A_{1g}$ and $E_g$ peaks monotonically blue shift with the decreasing of temperature. No CDW ordering is observed down to our measurement limit of 93 K, in strikingly contrasting to the CDW transition in bulk IrTe$_2$. From Fig. 2d, we find that those six peaks of area #2 merge into two $E_g$ and $A_{1g}$ peaks at 605 K, indicating that the $T_{CDW}$ is enhanced to 605 K. This value doubles the $T_{CDW}$ of the bulk IrTe$_2$. Similarly, the six-peak pattern is converted to two peaks by increasing the heating power of the incident laser (Fig. S5). The enhancement of $T_{CDW}$ has been reported in the bulk IrTe$_2$Se$_{2-x}$, where the lattice contractions of $a$ and $c$ increase the $T_{CDW}$ to 560 K.[20] Based on the relationship between $T_{CDW}$ and lattice constants of IrTe$_2$Se$_{2-x}$, we roughly estimate that the compressive in-plane strain generated in area #2 is about 2%. This is a large strain achieved in few-layer samples, which is approximately ten times stronger than that applied on IrTe$_2$ and Fe$_3$GeTe$_2$ thin flake.[20,37]

In Fig. 2e and 2f, the distinct phase boundary of areas #1 and #2 is clearly distinguished by the spatial mapping of Raman intensity of given Raman peaks. The structural symmetry is very important to understand the ground states, so we use the ARPRS to determine the space group of areas #1 and #2, see Fig.S6. It is found that the angle-dependent Raman intensity of areas #1 and #2 are identical to features of the high-$T$ and low-$T$ CDW phase of bulk IrTe$_2$, respectively. Therefore, the space groups at room temperature of areas #1 and #2 are $P$-3$m$1 and $P$-3, respectively.

We carried out the HRTEM measurements to investigate the origin of enhanced CDW ordering in area #2. The typical morphology of a large wrinkle with a height of ~110 nm is mapped out by the AFM (Fig. 3a). The transverse and longitudinal spans are 2.5 μm and 20 μm, respectively. We used the focused ion beam (FIB) to cut a slice of wrinkle together with the substrate and directly measured the atomic distance along the $c$-axis of the wrinkle. The fabricating steps are shown in Fig.S7 and Fig.S8. Figure 3b shows a typical cross-sectional TEM image of the wrinkle, which exhibits an arch-like morphology. As shown in the magnified TEM image in Fig. 3c and d, the interlayer distance, i. e. $c$-axis, is 5.26 Å. Meanwhile, from the high angle annular dark-field scanning transmission electron microscopy (HAADF-STEM), the lattice constant $a$ is 3.68 Å. the $c/a$ ratio of 1.43 significantly larger than the bulk values in IrTe$_{2-x}$Se$_x$ (1.37–1.40).[20] The enlarged $c/a$ ratio is used to explain the enhancement of $T_{CDW}$ from 280 K to 560 K in IrTe$_{2-x}$Se$_x$, in which the weaker interlayer coupling favors the higher $T_{CDW}$. Here the much enlarged $c/a$ ratio can perfectly certify the important role of weak interlayer interaction in enhancing $T_{CDW}$. Note that the CDW orders along a special direction of (10-1),[16] unfortunately, we do not observe the structural modulations from side view or top view.

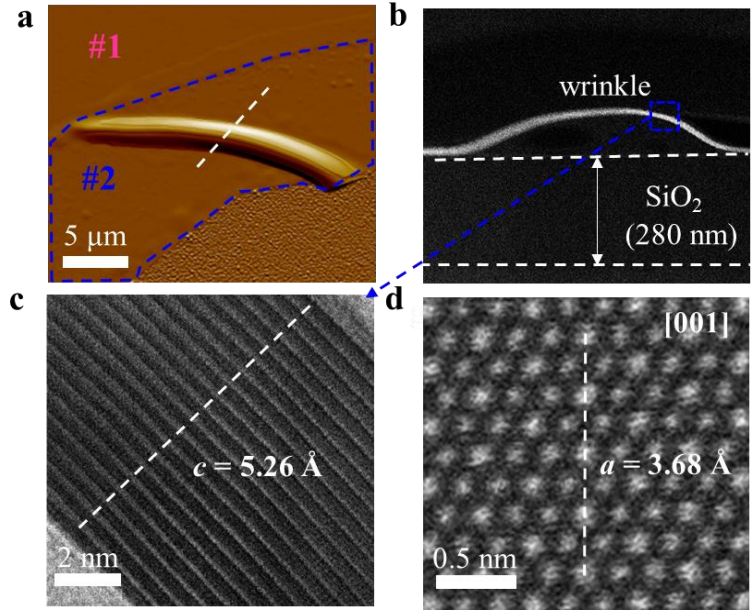

**Figure 3.** (a) AFM topography image of a typical wrinkle at the edge of IrTe$_2$ nanoflake. (b) Cross-sectional TEM image of the wrinkle on the SiO$_2$/Si substrate. (c) HRTEM image of the interlayer distance of the wrinkle marked by the blue square in **b**. (d) HAADF-STEM image of area #2 taken along the [001] axis.

We measure the intrinsic electrical transport properties of areas #1 and #2 to check the difference in the ground state. As illustrated in Fig. 4a, we intent to measure these two areas on the same IrTe$_2$ flake. Figure 4b shows the optical image of a sample. One can find that the flake in Fig. 4b is similar to the flake in Fig. 2a, where wrinkles can be clearly seen in area #2. Temperature dependence of the normalized resistances $R/R$(300K) from 1.5 to 300 K of are plotted in Fig. 4c and d with the optical image of the measuring device. We can see that area #1 shows a sharp superconducting transition at $T_c$=3.5 K, while area #2 remains metallic above 1.5 K. The superconducting transition can be gradually suppressed by the external magnetic field, and the estimated upper-critical-field $\mu_0 H_{c2}(0)$ through Ginzburg–Landau (GL) fitting is 0.15 T, see Fig. S9. Upon cooling the wrinkle-free samples, a clear SC of 3.5 K is obtained (Fig. S10).

It is known that dimensionality has profound effects on CDW instabilities, magnetism, and other phase transitions.[38,39,40] For 3D CDW, thinning the sample could destabilize the ordering kinetics along the layer-stacked direction and thus weaken the CDW order. In bulk IrTe$_2$, the superstructure vector is $q$ = 1/5(1, 0, -1) below $T_{CDW}$. The ordering kinetics of CDW slows with decreasing thickness in the $c$-direction,[28] resulting in the survival of the initial phase $P$-3$m$1. Besides, the enhanced fluctuations in the low-dimensional system lead to redistribution and disproportion of charge, mimicking the effect of carrier tuning. Therefore, reducing thickness, in turn, destroys the CDW ordering and induces SC under nearly equilibrium conditions.

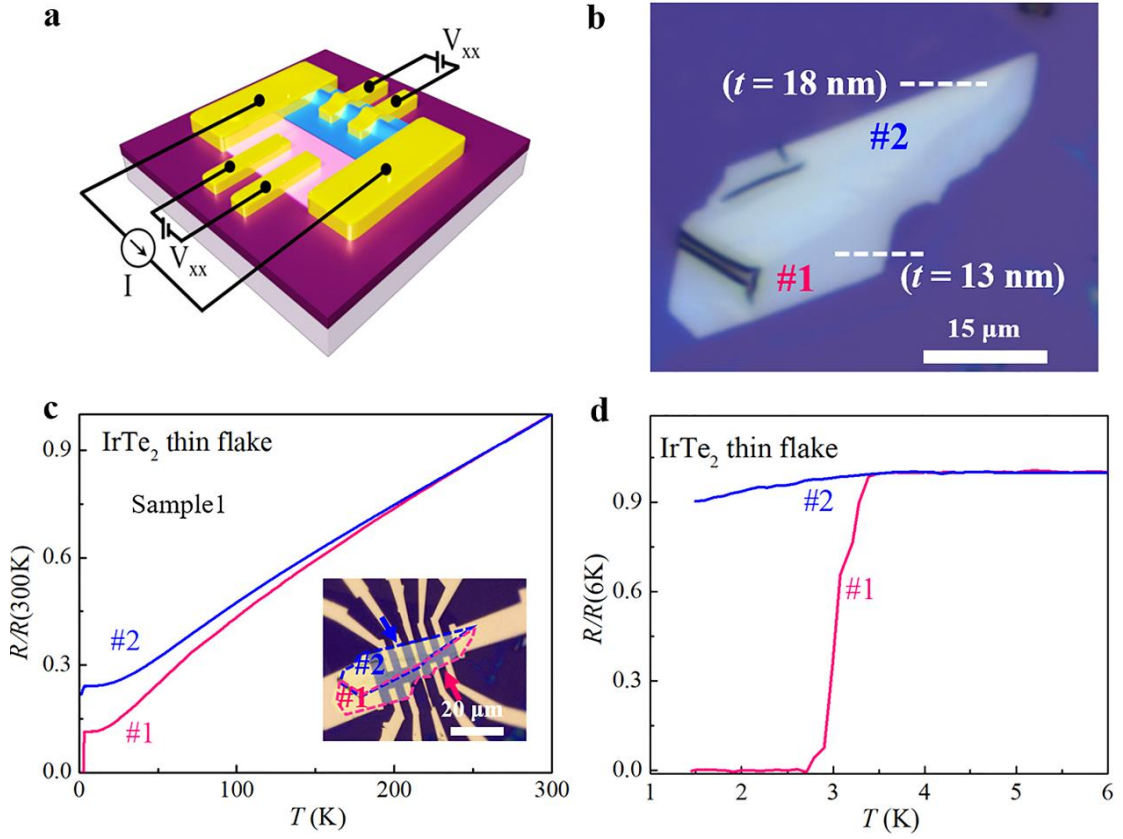

**Figure 4.** (a) Schematic diagram of the device fabricated on the $IrTe_2$ nanoflake with areas #1 and #2. Pink and blue regions represent areas #1 and #2, respectively. (b) Optical image of $IrTe_2$ nanoflake. (c) Temperature-dependent resistance ($R$) of areas #1 and #2. $R$ is normalized by the values at 300 K. Inset shows the optical image of the fabricated device. (d) $R(T)/R(6K)$ of two areas from 1.5 to 6 K.

In area #2, the situation is totally different. As documented in previous works, the bulk $IrTe_2$ favors the CDW ground state under in-plain tensile strain and Se-doping because the interlayer Te-Te bond becomes weak.[20,26] For example, exertion of 0.1% tensile-strain induces electron transfer from Ir to Te and destabilizes the Te-Te covalent bond, and thus enhancing the $T_{CDW}$ to 300 K. Besides, applying chemical pressure, i.e. Se-doping enlarges the $c/a$ ratio unexpectedly and results in the electron transfer from Se to Te because of the higher electronegativity of Se. It also weakens the Te-Te bond and enhances $T_{CDW}$ to 560 K. Thus, the subtle variations of electron number and microstructure in Te-Te bonds can significantly change the forming or melting kinetic of CDW order. As shown above, there are large wrinkles distributed at the edge or boundary of our nanoflake, the strain-field distributed may subtly alter the atomic distances. Based on the structural details of the wrinkles of graphene and other TMDC, the interlayer coupling is weakened by as large as 15%,[41,42] which modifies the electronic structure. In our work, the $c/a$ ratio of 1.43 around the wrinkle is significantly larger than 1.39 of $IrTe_{0.9}Se_{1.1}$ ($T_{CDW}$=560 K). It indicates that the interlayer Te-Te bonds are weakened, and the $T_{CDW}$ is enhanced much more. It is noted that this $c/a$ ratio is still smaller than the values (1.6-1.8) in common van der Waals materials.[43,44,45] Further tuning the *c*/*a* ratio is of interest to understand the intriguing relationship between chemical bonds and ground states.

The inconsistent observations about the coexistence, competition, and collaboration scenarios between SC and CDW in IrTe$_2$ come from the microscopic phase separation or CDW domain migrations in a thick sample. Relying on the conventional exfoliation method, it is hard to achieve either pure SC or pure CDW phase in a single piece of nanoflake because of its sensitive structure against physical properties. Taking advantage of a more advanced Au-assisted method, we successfully obtain thinner IrTe$_2$ nanoflake and provide solid evidence of competition pictures between SC and CDW. The pure SC phase and unusual enhancement of CDW ordering are attributed to the dimensional fluctuations and wrinkle-induced microstructure change, respectively. Our work reveals that the role of subtle structure in determining the ground states of IrTe$_2$, which might be extended to 2D materials to investigate the interplay of superconducting and other order parameters.

## ASSOCIATED CONTENT
**Supporting Information**
Methods of Au-assisted exfoliation process for cleaving IrTe$_2$ nanoflake; Temperature dependence of Raman spectrum for bulk IrTe$_2$; Laser power-dependent Raman spectra; SEM and TEM images of preparing steps for IrTe$_2$ nano-flake; R(T) curves of sample under external magnetic field.

## AUTHOR INFORMATION
**Corresponding authors:**
Yuan Huang, yhuang@bit.edu.cn;
Jian-gang Guo, jgguo@iphy.ac.cn;
Xiaolong Chen, xlchen@iphy.ac.cn
**Notes**
The authors declare no competing financial interest.

## ACKNOWLEDGMENTS
This work is supported by the National Key Research and Development Program of China (Grant Nos. 2019YFA0308000, 2017YFA0304700, 2018YFE0202601 and 2016YFA0300600), the National Natural Science Foundation of China (Grant Nos. 51922105, 11874405, 62022089 and 51772322), the Chinese Academy of Sciences (Grant No. QYZDJ-SSW-SLH013), the Beijing Natural Science Foundation (Grant No. Z200005).